\documentclass[12pt]{article}

\usepackage[margin=1in]{geometry}  	
\usepackage{amsfonts}
\usepackage{amsmath} 
\usepackage{booktabs}
\usepackage{multirow}
\usepackage{bbm}
\usepackage{subfig}
\usepackage{relsize}

\usepackage[doublespacing]{setspace} 
\usepackage{float}
\usepackage{caption}
\usepackage{graphicx}
\usepackage{adjustbox}
\usepackage{lscape}

\usepackage{xcolor}

\usepackage{tikz}
\usetikzlibrary{positioning,shapes.geometric}
\usetikzlibrary{arrows}

\usepackage{authblk}

\usepackage{cite}

\DeclareMathOperator{\E}{\mbox{E}}

\definecolor{forestgreen}{RGB}{34,139,34}
\usepackage{hyperref}
\hypersetup{
    colorlinks=true,
    linkcolor=magenta,
    filecolor=magenta, 
    citecolor=forestgreen,      
    urlcolor=blue
}

\usepackage[utf8]{inputenc}
\DeclareUnicodeCharacter{200E}{}

\usepackage{mathrsfs}

\usepackage{datetime}

\makeatletter
\def\@seccntformat#1{\@ifundefined{#1@cntformat}%
   {\csname the#1\endcsname\quad}  
   {\csname #1@cntformat\endcsname}
}
\let\oldappendix\appendix 
\renewcommand\appendix{%
    \oldappendix
    \newcommand{\section@cntformat}{\appendixname~\thesection\quad}
}
\makeatother

\usepackage[absolute,showboxes]{textpos}

\TPMargin{5pt}

\newcommand{\copyrightstatement}{
    \begin{textblock}{0.84}(0.08,0.93)    
         \noindent
         \footnotesize
         This draft manuscript presents work in progress. \\
         Comments and reports of mistakes are very much welcome at \href{mailto:issa\_dahabreh@brown.edu}{issa\_dahabreh@brown.edu}.
    \end{textblock}
}

\def\paperversionmajor{31}
\def\paperversionminor{0}

\begin{document}

\title{Study designs for extending causal inferences from a randomized trial to a target population}


\author[1,2,3,4]{Issa J. Dahabreh}
\author[5]{Sebastien J-P.A. Haneuse}
\author[4,5]{James M. Robins}
\author[1,2]{Sarah E. Robertson}
\author[6]{Ashley L. Buchanan}
\author[7]{Elisabeth A. Stuart}
\author[4,5]{Miguel A. Hern\'an}

\affil[1]{Center for Evidence Synthesis in Health, Brown University School of Public Health, Providence, RI}
\affil[2]{Department of Health Services, Policy \& Practice, School of Public Health, Brown University, Providence, RI}
\affil[3]{Department of Epidemiology, School of Public Health, Brown University, Providence, RI}
\affil[4]{Department of Epidemiology, Harvard T.H. Chan School of Public Health, Harvard University, Boston, MA}
\affil[5]{Department of Biostatistics, Harvard T.H. Chan School of Public Health, Harvard University, Boston, MA}
\affil[6]{Department of Pharmacy Practice, College of Pharmacy, University of Rhode Island, RI}
\affil[7]{Departments of Mental Health, Biostatistics, and Health Policy and Management, Johns Hopkins Bloomberg School of Public Health, Baltimore, MD}

\copyrightstatement

\maketitle{}
\thispagestyle{empty}

\clearpage
\pagenumbering{arabic}
\vspace*{0.8in}

\abstract{
\noindent
We examine study designs for extending (generalizing or transporting) causal inferences from a randomized trial to a target population. Specifically, we consider nested trial designs, where randomized individuals are nested within a sample from the target population, and non-nested trial designs, including composite dataset designs, where a randomized trial is combined with a separately obtained sample of non-randomized individuals from the target population. We show that the causal quantities that can be identified in each study design depend on what is known about the probability of sampling non-randomized individuals. For each study design, we examine identification of potential outcome means via the g-formula and inverse probability weighting. Last, we explore the implications of the sampling properties underlying the designs for the identification and estimation of the probability of trial participation.}

\clearpage


\noindent
A number of recent contributions \cite{cole2010, buchanan2018generalizing, dahabreh2018generalizing, westreich2017, lesko2017practical, rudolph2017, dahabreh2018transporting} have discussed methods for addressing problems related to selective study participation \cite{keiding2016perils} in randomized trials. These methods can be used to extend (i.e., generalize or transport \cite{hernan2016discussionkeiding}) causal inferences from a randomized trial to a target population. The methods require baseline covariate, treatment, and outcome data from individuals participating in the trial and baseline covariate data from non-randomized individuals. Estimation of potential outcome means in the target population typically requires models for the probability of trial participation \cite{cole2010}, the expectation of the outcome under each treatment among trial participants \cite{lesko2017practical}, or both (to improve robustness \cite{dahabreh2018generalizing, dahabreh2018transporting}). Prior work has largely focused on identifiability conditions and estimation approaches, without a clear connection to study design principles, obscuring the fact that different study designs determine which causal quantities can be identified and have implications for identifying and estimating the conditional probability of trial participation.

Two types of study designs that combine data from \emph{randomized individuals} with data from a sample of \emph{non-randomized individuals} have been used for the \emph{explicit goal} of estimating potential outcome means and treatment effects in a well-defined target population: (1) \emph{nested trial designs}, in which the randomized trial is embedded in a sample from the target population \cite{dahabreh2018generalizing}; and (2) \emph{non-nested trial designs}, in which observations from randomized individuals are combined with a separately obtained sample of non-randomized individuals from the target population. The sampling probability of non-randomized individuals is \emph{known} in nested trial designs \cite{buchanan2018generalizing}, but \emph{unknown} in non-nested trial designs \cite{westreich2017,rudolph2017, dahabreh2018transporting}. In both types of study designs, baseline covariate data are collected from all randomized individuals and from sampled non-randomized individuals; treatment and outcome data need only be collected from randomized individuals. Though treatment and outcome data from non-randomized individuals can be used to evaluate assumptions or improve efficiency, they are not necessary for identification and estimation under the assumptions used in this paper or the bulk of the related literature (e.g., as reviewed in \cite{lesko2017practical}).

In this paper, we show that the different causal quantities that can be identified in each study design depend on what is known about the probability of sampling non-randomized individuals. For each study design, we examine identification of potential outcome means via the g-formula and inverse probability weighting, and explore the implications of the design's sampling properties for modeling the probability of trial participation.

\section{Sampling properties and the observed data}\label{sec:sampling_properties}

For a well-defined causal question, investigators can specify a set of eligibility criteria that define an \emph{actual population} of individuals to whom research findings would be applicable, in the sense that we can \emph{in principle} identify all individuals who meet the criteria. For instance, when designing a randomized trial, the trial eligibility criteria define an actual population of all trial-eligible individuals. In this paper, we view the actual population as a simple random sample from an (infinite) \emph{superpopulation} of individuals; we refer to this superpopulation as the \emph{target population} \cite{robins1988confidence}. We are interested in causal quantities that pertain to the target population or to its subsets (e.g., defined by trial participation status).

To introduce some notation, let $X = (X_1, \ldots, X_p)$ denote a vector of $p$ baseline covariates; $A$ the treatment assignment indicator; $Y$ the observed outcome; and $S$ the trial participation indicator, with $S=1$ for randomized individuals and $S=0$ for non-randomized individuals (individuals who are either not invited to participate in the trial or who are invited but decline). To capture the notion that some non-randomized individuals in the actual population ($S=0$) may not be sampled, let $D$ be an indicator for whether an individual in the actual population is sampled and contributes data to the analyses, with $D=1$ for sampled individuals and $D=0$ for non-sampled individuals.. 

\begin{figure}[ht!]
\caption{Conceptual graph depicting the sampling designs for studies extending inferences from a randomized trial to a target population.}\label{figure_1}
	\centering
\includegraphics[scale=0.85]{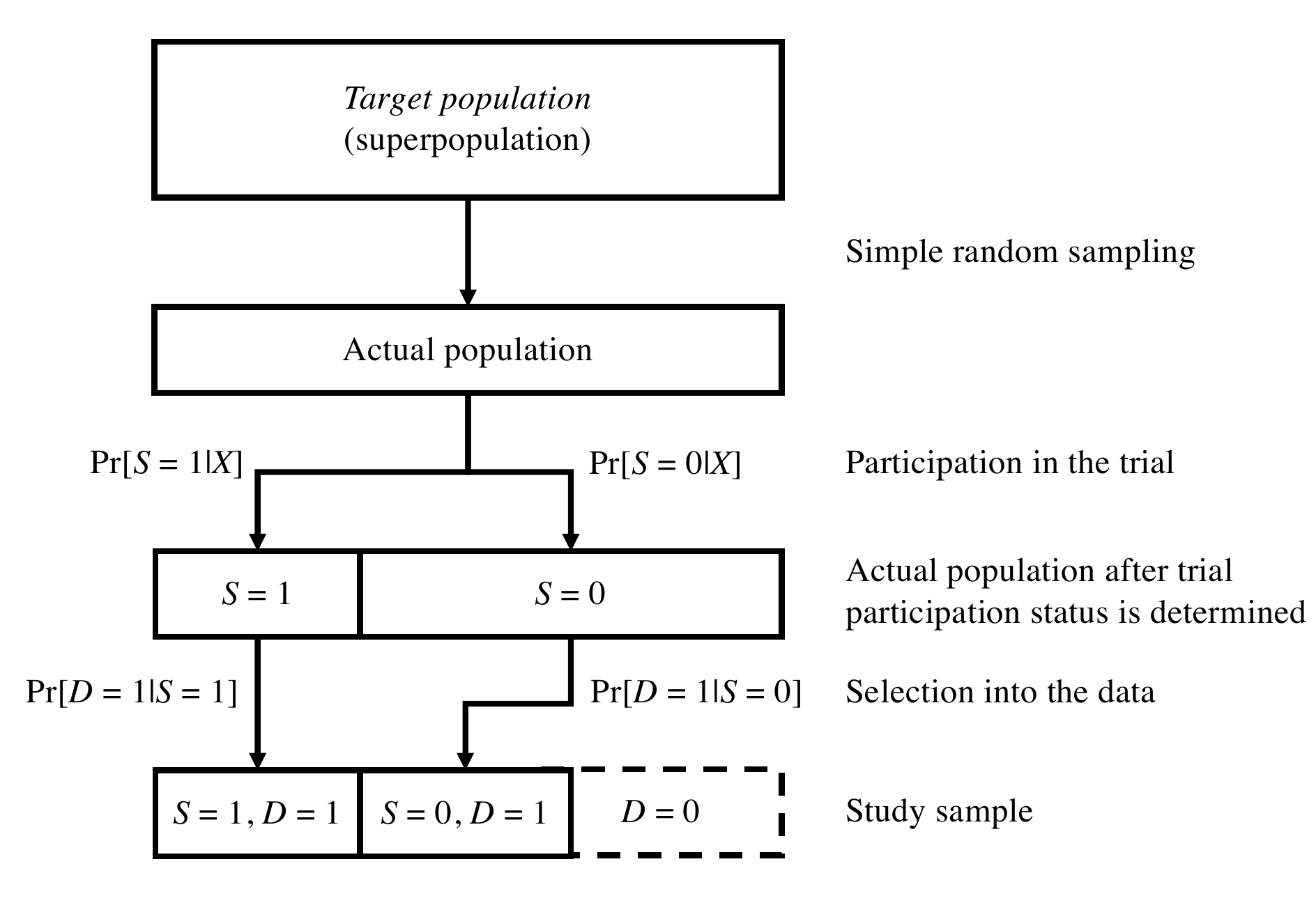}
\end{figure}

We can now describe the \emph{sampling properties} that underlie nested and non-nested study designs. These properties describe how the observed study sample relates to the actual population; the underlying actual population and (hypothetical) target population are the same. Figure \ref{figure_1} illustrates the conceptual relationships between designs, their sampling properties, and the observed data.

In the main text of this paper, we consider simple random samples, with \emph{known or unknown sampling probabilities}, from the actual population or from the non-randomized subset of the actual population. As we discuss below, our main results, with minor modifications, hold when the sampling probability is a known function of auxiliary baseline covariates rather than a known constant (i.e., when we have random sampling, not \emph{simple} random sampling). Allowing the sampling probabilities to depend on auxiliary covariates, however, does not lead to additional insights regarding study design \cite{dahabreh2019generalizing}; for this reason, in the main text, we assume that the sampling probability does not depend on covariates.

\subsection{Nested trial designs}\label{subsec:nested_trials_sampling_prop}

We consider two variants of the nested trials design: when a \emph{census of the actual population} is taken and when the \emph{non-randomized individuals are sub-sampled}.

\emph{Census of the actual population:} In this variant of the nested trial design, the individuals contributing data to the analysis are assumed to be a census of the actual population, that is, 
\begin{equation*}
		\Pr[D = 1 | S = 1] = \Pr[D = 1 | S = 0] = 1,
\end{equation*}
thus, nested trial designs can be viewed as simple random samples from the superpopulation. In this design, it is common to define the target population implicitly, based on the actual population in which the trial is nested. For example, in comprehensive cohort studies \cite{Olschewski1985}, investigators nest a trial within a cohort of all individuals who met the trial's eligibility criteria and were invited to participate in the trial. They then define the target population as the population from which cohort members (i.e., the actual population of trial-eligible individuals invited to participate in the trials) could have been a simple random sample. Thus, in this design, investigators need to ensure that the cohort represents the target population they are interested in; that is, the trial eligibility criteria need to be broad enough to address the research question \emph{and} the individuals invited to participate in the trial (who form the cohort in which the trial is nested) need to represent the target population of interest.

\emph{Sub-sampling of non-randomized individuals:} In this variant,  we collect data from all randomized individuals in the actual population but only collect baseline covariate data from a sub-sample of the non-randomized individuals in the actual population, with sampling probability that is a \emph{known} constant. The sampling properties can be summarized as 
\begin{equation*} \label{eq:sampling_ass_2sample}
	\begin{split}
		&\Pr[D = 1 | S = 1] = 1 \mbox{ and } \\
    &\Pr[D = 1 | X, A, Y, S = 0] = \Pr[D = 1 | S = 0]= c > 0,
	\end{split}
\end{equation*}
where $c$ is a known constant, with $0 < c \leq 1$. Note that the nested trial design with a census of the actual population can be viewed as a special case of the sub-sampling design, with $c=1$. Using $c < 1$ is statistically less efficient than using $c=1$, but may improve research economy, for example, if the collection of covariate data among non-randomized individuals is expensive \cite{dahabreh2019generalizing}. Furthermore, as noted, a variant of the nested trial design with sub-sampling allows the selection of non-randomized individuals to depend on auxiliary baseline covariates; we show how our results extend to that case in the Appendix.

\subsection{Non-nested trial designs}\label{subsec:non_nested_trials_sampling_prop}

In non-nested trial designs, data from randomized and non-randomized individuals are obtained separately. Investigators assume that data from all randomized individuals can be combined with data from a simple random sample of non-randomized individuals from the actual population, with sampling probability that was an \emph{unknown} constant (e.g., \cite{westreich2017}). The sampling properties can be summarized as
\begin{equation*} \label{eq:sampling_ass_2sample}
  \begin{split}
		&\Pr[D = 1 | S = 1] = 1 \mbox{ and } \\
		&\Pr[D = 1 | X, A, Y, S = 0] = \Pr[D = 1 | S = 0] = u > 0, \mbox{ where $u$ is an unknown constant.}
  \end{split}
\end{equation*}

An example of non-nested trial design is the \emph{composite dataset design} \cite{westreich2017, dahabreh2018transporting}. Here, investigators append the data from a randomized trial to data from a convenience sample of non-randomized individuals, often obtained from routinely collected data sources (such as claims or electronic medical records databases, or prospective cohort studies). The assumption, often left unstated, is that the sample of non-randomized individuals is a simple random sample from the population of non-randomized individuals (or a well-defined subset thereof) to whom the investigators wish to extend the trial results \cite{dahabreh2018transporting, westreich2017}.

\subsection{The observed data}
In both nested and non-nested designs, we collect data on baseline covariates, treatment, and outcome from randomized individuals; in contrast, as we show in Section \ref{sec:identification_g_form}, only baseline covariate data are needed from non-randomized individuals. 

More specifically, for nested designs the observed data consist of realizations of 
\begin{equation*}
  \begin{split} 
&(X, A, Y, S = 1, D = 1) \mbox{ for trial participants;} \\
&(X, S = 0, D = 1) \mbox{  for \emph{sampled} non-randomized individuals;} \\
&(S = 0, D = 0) \mbox{  for \emph{non-sampled} non-randomized individuals.}
  \end{split}
\end{equation*}
Because all randomized individuals are sampled, we have that $(D = 1, S = 1) \Leftrightarrow (S = 1)$. No covariate, treatment, or outcome data are available for non-sampled non-randomized individuals, $(D = 0, S = 0)$. Note also that in nested trial designs with a census of the actual population, the $(D = 0, S = 0)$ subset does not exist.

In non-nested trial designs (e.g., composite dataset designs), we typically do not know the number of non-sampled non-randomized individuals, thus the observed data consist of realizations only of
\begin{equation*}
  \begin{split} 
&(X, A, Y, S = 1, D = 1) \mbox{ for trial participants; and} \\
&(X, S = 0, D = 1) \mbox{  for \emph{sampled} non-randomized individuals.} 
  \end{split}
\end{equation*}

\section{Causal quantities of interest and identifiability conditions}\label{sec:identifiability_conditions}

\subsection{Causal quantities of interest}

In order to define causal quantities, let $Y^a$ be the potential (counterfactual) outcome under intervention to set treatment to $a$ \cite{rubin1974, robins2000d}. We are interested in the mean of these potential outcomes in the target population $ \E[Y^a] $ or in the non-randomized subset of the target population $ \E[Y^a | S = 0] $. For example, $ \E[Y^a] $ captures the outcome under the strategy of treating all individuals in the target population with $a$. And it is often scientifically and methodologically interesting to compare $ \E[Y^a | S = 0] $ against $ \E[Y^a | S = 1] $, to examine whether the potential outcome mean under treatment $a$ differs among trial-participants and non-participants in the target population \cite{dahabreh2018generalizing}.

\subsection{Identifiability conditions}

For all study designs, the following identifiability conditions are sufficient to extend inferences from a clinical trial to a target population \cite{dahabreh2018generalizing, dahabreh2018transporting}:

\noindent
(1) \emph{Consistency of potential outcomes:} interventions are well-defined, so that if $A_i = a,$ then $Y^a_i = Y_i$. Implicit in this notation is that the offer to participate in the trial and trial participation itself do not have an effect on the outcome except through treatment assignment (e.g., there are no Hawthorne effects).

\noindent
(2) \emph{Mean exchangeability among trial participants:} $\E [ Y^a | X, S = 1, A = a] = \E [ Y^a | X, S = 1] $. This condition is expected to hold because of randomization (marginal or conditional on $X$). 

\noindent
(3) \emph{Positivity of treatment assignment in the trial:} $\Pr[A =a | X = x, S = 1] > 0$ for each $a$ and each $x$ with positive density among randomized individuals, $f_{X|S}(x | S = 1) > 0$. This condition is also expected to hold because of randomization.

\noindent
(4) \emph{Mean generalizability (exchangeability over $S$):} $\E[ Y^a | X, S = 1] = \E[Y^a | X]$ for each $a$. For binary $S$, this condition implies the mean transportability condition $\E[ Y^a | X, S = 1] = \E[Y^a | X, S = 0 ]$ (provided both conditional expectations are well-defined).

\noindent
(5) \emph{Positivity of trial participation:} $\Pr[S = 1 | X = x] > 0$ for each $x$ with positive density in the target population, $f_X(x) > 0$.

In the conditions listed above, we have used $X$ generically to denote baseline covariates. It is possible, however, that strict subsets of $X$ are adequate to satisfy different exchangeability conditions. For example, in a marginally randomized trial the mean exchangeability among trial participants holds unconditionally. Furthermore, to focus on issues related to selective trial participation, we will assume complete adherence to the assigned treatment and no loss-to-follow-up.

\subsection{Trial eligibility criteria and choice of target population}

Now that we have specified the causal quantities of interest and listed identifiability conditions, we can consider the choice of target population in more detail. As noted, the target population should be determined by the scientific question investigators hope to address. In many cases, when using the methods described in this paper, it is sensible to limit the target population to the population of individuals meeting the trial eligibility criteria or to a subset of that population. To the extent that the variables used to define the trial eligibility criteria are needed to satisfy the mean generalizability condition, the restriction to trial-eligible individuals is needed for the positivity of trial participation condition to hold -- individuals not meeting the criteria are not allowed to enter the trial. In some cases, however, investigators may be able to argue that only a subset of the variables used to determine trial eligibility are necessary for the mean generalizability condition to hold. In such cases, the target population can be broader than the population of trial eligible individuals. The essential requirement is that the distributions of covariates needed to satisfy the mean generalizability condition among randomized and non-randomized individuals should have common support.

\section{Identification via the g-formula}\label{sec:identification_g_form}

We begin by considering identification by the g-formula \cite{robins1986}. Using the identifiability conditions of Section \ref{sec:identifiability_conditions}, it is straightforward to show that the potential outcome mean in the target population \cite{dahabreh2018generalizing} can be re-expressed as 
\begin{equation} \label{eq:idmain}
	\begin{split}
\E[Y^a] 	&= \E\! \big[ \E[Y | X, S = 1 , A = a] \big] \\
		&= \int \E[Y | X = x, S = 1 , A = a] dF_{X}(x),
	\end{split}
\end{equation}
where $F_{X}(x)$ denotes the distribution of $X$ in the target population.

The potential outcome mean among non-randomized individuals in the target population \cite{dahabreh2018transporting} can be re-expressed as
\begin{equation} \label{eq:idmainS0}
	\begin{split} 
\E[Y^a | S = 0] 	&= \E\! \big[ \E[Y | X, S = 1 , A = a] | S = 0 \big] \\
				&= \int \E[Y | X = x, S = 1 , A = a] dF_{X|S}(x | S = 0),
	\end{split}
\end{equation}
where $F_{X|S}(x | S = 0)$ denotes the distribution of $X$ among non-randomized individuals in the target population (i.e., the subset with $S = 0$).

First, we note that both results involve the conditional expectation of the outcome $Y$ among trial participants assigned to treatment $a$, $\E[Y| X, A= a, S = 1]$. Because both nested and non-nested designs assume that all randomized individuals are sampled, this expectation is identifiable in both designs. 

Next, we turn out attention to the identification of $F_X(x)$ and $F_{X|S}(x|S=0)$, which are necessary to identify $\E[Y^a]$ and $\E[Y^a | S = 0]$, respectively. There are interesting differences between the designs when it comes to identifying these distributions and we consider each design individually below.

\subsection{Nested trial designs}

\emph{Census of the actual population:} Identification is most straightforward in this design, because data are available from all members of the actual population (both randomized and non-randomized) and the actual population is a simple random sample from the target population. Thus, $F_{X}(x)$ is identifiable. Furthermore, in this design, every subgroup of the actual population defined on the basis of baseline covariates or trial participation is a simple random sample from the corresponding subgroup in the target population. Thus, the distribution of covariates among non-randomized individuals $F_{X|S}(x | S = 0)$ can also be identified. It follows that all the components on the right-hand-sides of (\ref{eq:idmain}) and (\ref{eq:idmainS0}) are identifiable, establishing that $\E\! \big[ \E[Y | X, S = 1 , A = a] \big]$ and $\E\! \big[ \E[Y | X, S = 1 , A = a] \big| S = 0 \big]$ are identifiable.

\emph{Sub-sampling of non-randomized individuals:} For this design, identification of the marginal distribution of $X$ is slightly more involved because the non-randomized individuals contributing data to the analysis are a sub-sample from the non-randomized individuals in the actual population. 

First, by the law of total probability, for binary $S$, $$F_X(x) = \sum\limits_{s = 0}^{1} F_{X|S}(x | S = s) \Pr[S = s].$$ Clearly, $F_{X|S}(x | S = s)$, for $s = 0, 1$ is identifiable because the randomized and non-randomized sampled individuals are simple random samples of the target population subsets with $S=1$ and $S=0$, respectively. The only difficulty, then, is identification of the marginal probability of trial participation, $\Pr[S = 1]$, because $\Pr[S = 0] = 1 - \Pr[S = 1]$. As we show in the Appendix, under the sampling properties of the nested trial design with sub-sampling of non-randomized individuals,
\begin{equation} \label{eq:marginal_prob}
\Pr[S = 1] = \left\{1 + \dfrac{\Pr[S = 0 | D = 1]}{\Pr[S = 1 | D = 1]} \times c^{-1}   \right\}^{-1}.
\end{equation}
The odds of non-participation in the trial among sampled individuals, $ \dfrac{ \Pr[S = 0 | D = 1] }{ \Pr[S = 1 | D = 1]}$, are identifiable; and, as defined in Section \ref{subsec:nested_trials_sampling_prop}, $c$ is a known constant. It follows that $F_X(x)$ is identifiable and, consequently, $\E\! \big[ \E [Y | X, S = 1, A = a]\big]$ is also identifiable because all the components of the integral on the right-hand-side of (\ref{eq:idmain}) are identifiable. 

Turning our attention to $F_{X|S}(x | S = 0)$, we note that it is identifiable because the sampled non-randomized individuals are a simple random sample from the non-randomized individuals in the actual population. It follows that $\E\! \big[\! \E[Y | X, S = 1, A = a] | S = 0 \big]$ is identifiable because all the components of the integral on the right-hand-side of (\ref{eq:idmainS0}) are identifiable.

\subsection{Non-nested trial designs}

Using an argument parallel to that for nested trial designs with sub-sampling, when the probability of sampling a non-randomized individual is unknown, the probability of trial participation, $\Pr[S = 1]$, can be expressed in the form of equation (\ref{eq:marginal_prob}), substituting the $u$ for $c$,
\begin{equation*} 
\Pr[S = 1] = \left\{1 + \dfrac{\Pr[S = 0 | D = 1]}{\Pr[S = 1 | D = 1]} \times  u^{-1} \right\}^{-1}.
\end{equation*}
Because, as defined in Section \ref{subsec:non_nested_trials_sampling_prop}, $u$ is an unknown constant, $F_{X}(x)$ is not identifiable and consequently $\E\! \big[ \E[Y | X, S = 1 , A = a] \big]$ is also not identifiable.

Turning our attention to $F_{X|S}(x | S = 0)$, we see that it is identifiable because the non-randomized individuals contributing data to the analysis are a simple random sample from the non-randomized individuals in the actual population (even though the sampling probability is unknown). It follows that $\E\! \big[ \E[Y | X, S = 1 , A = a] \big | S = 0 \big]$ is identifiable in non-nested trial designs because all the components of the integral in (\ref{eq:idmainS0}) are identifiable.

\section{Identification via IP weighting}

There has been considerable recent interest \cite{cole2010, dahabreh2018generalizing, buchanan2018generalizing, westreich2017, dahabreh2018transporting} in using weighting methods to identify the potential outcome means in equations (\ref{eq:idmain}) and (\ref{eq:idmainS0}), because the specification of models for the probability of trial participation is often considered a somewhat easier task than the specification of models for the outcome among trial participants.

First, consider $\E\! \big[ \E[Y | X, S = 1 , A = a] \big]$, which we argued is identifiable in nested trials. As shown in \cite{dahabreh2018generalizing}, we can re-express the right-hand-side of (\ref{eq:idmain}) as 
\begin{equation}\label{eq:IP_id_prob}
	\begin{split}
	\E\! \big[ \E [Y | X, S = 1 , A = a]  \big]
	&= \E\! \left[ \dfrac{I(S = 1, A = a) Y }{\Pr[S = 1| X] \Pr[A = a | X, S = 1]} \right],
	\end{split}
\end{equation}
where $I(\cdot)$ denotes the indicator function.

Now, consider $\E\! \big[ \E [Y | X, S = 1 , A = a] | S = 0 \big]$, which we argued is identifiable by the g-formula in both nested and non-nested trials. As shown in \cite{dahabreh2018transporting}, we can re-express  the right-hand-side of (\ref{eq:idmainS0}) as 
\begin{equation}\label{eq:IP_id_odds}
	\begin{split}
	\E\! \big[ \E [Y | X, S = 1 , A = a] \big| S = 0  \big] 
  &= \dfrac{\quad \E\! \left[ \dfrac{I(S = 1, A = a) Y \Pr[S = 0| X] }{\Pr[S = 1| X] \Pr[A = a | X, S = 1]    }   \right]\quad  }{\E\! \left[ \dfrac{I(S = 1, A = a) \Pr[S = 0| X]}{\Pr[S = 1| X] \Pr[A = a | X, S = 1]} \right]}.
	\end{split}
\end{equation}

The probability of treatment among trial participants, $\Pr[A = a | X, S = 1]$ is under the control of the investigators and does not pose any difficulties for identification of either functional. Now, for each design, we focus our attention on the conditional probability or the conditional odds of trial participation, which appear in expressions (\ref{eq:IP_id_prob}) and (\ref{eq:IP_id_odds}), respectively.

\subsection{Nested trial designs}

\emph{Census of the actual population:} Identification of $\Pr[S = 1 | X]$ in this design is an obvious consequence of the fact that individuals contributing data to the analysis are a simple random sample from the target population. In other words, because we have sampled all individuals in the actual population, which is a simple random sample of the target population, $\Pr[S = 1 | X] = \Pr[S = 1 | X, D = 1]$.

\emph{Sub-sampling of non-randomized individuals:} Identification of $\Pr[S = 1 | X]$ is only a little more difficult when we sample non-randomized individuals from the actual population. As we show in the Appendix, under the sampling properties of this design, 
\begin{equation}\label{eq:odds1}
  \begin{split}
\Pr[S = 1 | X] =  \left\{ 1 + \dfrac{\Pr[S=0 | X , D = 1]}{\Pr[S = 1 | X, D = 1]} \times c^{-1} \right\}^{-1},
  \end{split}
\end{equation}
where the conditional odds of trial participation among sampled individuals, $\dfrac{\Pr[S=0 | X , D = 1]}{\Pr[S = 1 | X, D = 1]}$, are identifiable and $c$ is a known constant defined in Section \ref{subsec:nested_trials_sampling_prop}. It follows that $\Pr[S = 1 | X]$ is identifiable and the odds of trial participation can be written as
\begin{equation}\label{eq:odds2}
\dfrac{\Pr[S = 1| X]}{\Pr[S = 0 | X]} = \dfrac{\Pr[S = 1| X , D = 1]}{\Pr[S = 0 | X , D = 1]} \times c.
\end{equation}
In sum, the IP weighting re-expressions of the functionals of interest are identifiable in nested trial designs.

\subsection{Non-nested trial designs}

We can use an argument parallel to that for nested trial designs with sub-sampling, to establish that, when the sampling probability for non-randomized individuals is unknown, the probability of trial participation, $\Pr[S = 1 | X]$, can be expressed as,
\begin{equation}\label{eq:odds_u}
\Pr[S = 1 | X] =  \left\{ 1 + \dfrac{\Pr[S=0 | X , D = 1]}{\Pr[S = 1 | X, S = 1]} \times u^{-1} \right\}^{-1}.
\end{equation}
Because, as defined in Section \ref{subsec:non_nested_trials_sampling_prop}, $u$ is unknown, the conditional probability of trial participation, which appears on the right hand side of (\ref{eq:IP_id_prob}), is not identifiable; this confirms our earlier result that $\E\! \big[ \E [Y | X, S = 1 , A = a] \big]$ cannot be identified in non-nested trials. 

Furthermore, the conditional odds of trial participation are also not identifiable because they depend on $u$. In fact, using equation (\ref{eq:odds2}), substituting $u$ for $c$, we see that the odds of trial participation in the target population are, \emph{up to an unknown multiplicative constant}, equal to the odds of trial participation among sampled individuals,
\begin{equation}\label{eq:composite_odds}
\dfrac{\Pr[S = 1| X]}{\Pr[S = 0 | X]} = \dfrac{\Pr[S = 1| X , D = 1]}{\Pr[S = 0 | X , D = 1]} \times u.
\end{equation}

We have come to an apparent conflict: the right hand-side of (\ref{eq:IP_id_odds}) involves the conditional odds of trial participation, a quantity that is not identifiable in non-nested designs. Yet, we argued in the previous section that the left-hand-side of (\ref{eq:IP_id_odds}) is identifiable. The conflict can be easily resolved by noting that, because both the numerator and the denominator of (\ref{eq:IP_id_odds}) are multiplied by the unknown constant $u$, which cancels out, identification via IP weighting is possible (see the appendix of \cite{dahabreh2018transporting} for technical details).

Table \ref{tab:summary_conditions_identification} summarizes the sampling properties and identification results for each study design.

\section{Estimating the probability of trial participation}

In realistic analyses, the dimension of $X$ will be fairly large, necessitating some modeling assumptions about $\Pr[S=1|X]$ or $\Pr[S=1|X, D=1]$ \cite{robins1997toward}. In this section we discuss the relationship between study design and model specification and estimation approaches.

\subsection{Nested trial designs}

\emph{Census of the actual population:} In this type of nested trial design, it is straightforward to estimate the probability of trial participation, $\Pr[S=1|X]$, in the sense that we can use the model we believe is most likely to be correctly specified for the target population. 

For concreteness, suppose that we are willing to assume a parametric model, $p(X; \gamma),$for the probability of trial participation in the target population, $ \Pr[S = 1 | X]$, with $\gamma$ a finite dimensional parameter. In the nested-trial designs with a census of non-randomized individuals, we typically estimate the parameters by maximizing the likelihood function
\begin{equation*}
	\begin{split}
	\mathscr{L}(\gamma) &= \prod\limits_{i=1}^{n} \left[ p(X_i; \gamma)\right]^{S_i} \left[ 1 - p(X_i; \gamma)\right]^{1 - S_i},
	\end{split}
\end{equation*}
where $i=1,\ldots,n$, and $n$ is the number of individuals in the study (i.e., the actual population). Under reasonable technical conditions \cite{newey1994large}, the usual maximum likelihood methods use a sample-size normalized objective function that converges uniformly in probability to
\begin{equation}\label{eq:population_est}
	\ell_0(\gamma) = \E\! \Big[ S \log \big[ p(X; \gamma) \big] + (1 - S) \log \big[1 - p(X; \gamma)\big] \Big].
\end{equation}
For example, when $p(X; \gamma)$ is a logistic model, $\ell_0(\gamma)$ is the large sample limit of the sample-size normalized log-likelihood function for logistic regression.

\emph{Sub-sampling of non-randomized individuals:} When we sub-sample of the non-randomized individuals in the actual population, it is not possible to maximize the likelihood function above, because data are not available from all non-randomized individuals in the actual population. A natural idea is to use equation (\ref{eq:odds1}), which provides an explicit formula for identifying the conditional probability of trial participation, $\Pr[S=1|X]$ using the probability of trial participation among sampled individuals, $\Pr[S=1|X, D=1]$, and the sampling probability for non-randomized individuals, $\Pr[D=1|S=0]$. When modeling the probability of trial participation among sampled individuals, however, the following difficulty arises: in general, \emph{when sampling depends on trial participation status}, the correctly specified model for trial participation does not have the same form as the correctly specified model in the target population, with the notable exception of the logistic regression model \cite{manski1977estimation, scott1986fitting}. This implies that naive estimation of the parameters of the model for trial participation among sampled individuals will typically be inconsistent for the population model.

Because the sampling probability of non-randomized individuals is known, we can use the following weighted pseudo-likelihood function, which only uses data from sampled individuals \cite{manski1977estimation, cosslett1981maximum},
\begin{equation*}
  \begin{split}
  \mathscr{L}_{\scalebox{.5}{\text{W}}}(\gamma) &= \prod\limits_{i=1}^{n} \big[ p(X_i; \gamma)\big]^{S_i D_i} \big[ 1 - p(X_i; \gamma)\big]^{[ (1 - S_i) D_i ] / c},
  \end{split}
\end{equation*}
with $c=\Pr[D=1|S=0]$. Weighted maximum likelihood methods use a sample-size normalized objective function that converges uniformly in probability to
\begin{equation}\label{eq:population_est_weights}
  \ell_{\scalebox{.5}{\text{W}}0}(\gamma) = \E\! \left[ S D \log \big[ p(X; \gamma) \big] + \dfrac{[(1 - S) D}{c} \log \big[1 - p(X; \gamma)\big] \right],
\end{equation}
which is restricted to sampled individuals ($D = 1$).

As we show in the Appendix, under the sampling properties for this design, the large sample limits of the objective functions in (\ref{eq:population_est}) and (\ref{eq:population_est_weights}) are equal, $\ell_0(\gamma) =  \ell_{\scalebox{.5}{\text{W}}0}(\gamma)$. It follows that, under reasonable technical conditions \cite{newey1994large}, weighted likelihood estimation of $\gamma$ in the nested trial design with sub-sampling of non-randomized individuals converges in probability to the same parameter as unweighted regression in the actual population.

In practical terms, as long as a reasonable parametric model can be specified for the target population, the model parameters can be estimated using weighted maximum likelihood methods \cite{manski1977estimation} on data from sampled individuals, with individual level weights equal to 1 for randomized individuals, $S = 1, D = 1$; $c^{-1}$ for sampled non-randomized individuals, $S = 0, D = 1$; and 0 for unsampled individuals, $D = 0$.

\subsection{Non-nested trial designs}

In non-nested trial designs, the weighting approach described above is not applicable because the sampling probability of non-randomized individuals is unknown. Provided, however, that the sampling probability does not depend on $X$ (i.e., the assumed sampling property), we can show that, if a logistic model for trial participation is correctly specified in the target population, then a logistic model is correctly specified in the non-nested trial design. To see this, suppose that we are willing to assume a logistic regression model in the population, such that 
\begin{equation*}
\ln \dfrac{\Pr[S =1 | X ]}{\Pr[S = 0 | X]} =  \beta_0 + \sum\limits_{j = 1}^{p} \beta_j X_j.
\end{equation*}
Using the result in (\ref{eq:odds_u}) and taking logarithms, we have that 
\begin{equation*}
	\begin{split}
\ln \dfrac{\Pr[S =1 | X ]}{\Pr[S = 0 | X]} &= \ln(u) + \ln \dfrac{\Pr[S =1 | X , D = 1]}{\Pr[S = 0 | X, D = 1]}.
	\end{split}
\end{equation*}

Equating the right-hand-sides of the last two equations, we obtain
\begin{equation}\label{eq:logistic_model_subsample}
	\begin{split}
\ln \dfrac{\Pr[S =1 | X , D = 1]}{\Pr[S = 0 | X, D = 1]} 
		 &= \beta_0^* + \sum\limits_{j = 1}^{p} \beta_j X_j,
	\end{split}
\end{equation}
where $\beta_0^* = \beta_0 - \ln(u)$, a well-known result in the context of case-referent studies \cite{mantel1973synthetic}. Thus, if a logistic model is correctly specified in the target population, then a model of the same functional form is correctly specified in the non-nested trial design. In fact, the coefficients in the two models are equal, and only the intercept differs. Because $0 < u < 1$, $\beta_0^* > \beta_0$: the sub-sampling of non-randomized patients simply results in an intercept that is ``shifted'' upwards. As we have shown in the section on IP weighting, the resulting shift in the odds of participation does not affect identification of the potential outcome mean in the non-randomized individuals, $\E[Y^a | S = 0]$, which is the parameter of interest in non-nested trial designs with unknown sampling probability of non-randomized individuals.

The above result is also important for estimation of the model parameters: combined with the results in \cite{prentice1979logistic, breslow2000semi}, it implies that the unconstrained and unweighted maximum likelihood estimator for the logistic model in (\ref{eq:logistic_model_subsample}), fit among sampled individuals, is the efficient estimator for $\beta_j$, $ j = 1, \ldots, p$.

\section{Discussion}

We presented a unified description of study designs for extending inferences from randomized trials to a well-defined target population and showed that commonly invoked identifiability conditions need to be combined with the sampling properties of each study design in order to determine which causal quantities can be identified. Our approach uses a superpopulation framework, which is a natural choice for extending trial findings beyond the sample of randomized individuals \cite{hernan2019}. 

In non-nested trial designs, where the sampling probability for non-randomized individuals is \emph{unknown}, the marginal potential outcome means in the target population are not identifiable, but the potential outcome means in the sub-population of non-randomized individuals are identifiable. This restriction may be less severe than it appears: for most trials, we want to estimate the effect of applying the interventions to a new population, which can be represented by a well-chosen sample of non-randomized individuals \cite{dahabreh2018transporting}. In any case, when available, knowledge of the sampling probability of non-randomized individuals can be used to mitigate these limitations, without requiring the collection of covariate information from all non-randomized individuals in the actual population. Thus, in general, nested trial designs will often be the preferred approach for generalizing trial findings when it is possible to define and sample the actual population when a randomized trial is planned. Such nested trial designs will typically have broad (pragmatic \cite{ford2016}) eligibility criteria and define the target population as the population of individuals meeting the trial eligibility criteria. When that is not possible, as is the case in already completed randomized trials, non-nested trial designs might be a reasonable alternative. For example, in non-nested trial designs, the comparison of estimates for the potential outcome means among randomized, $\widehat \E[Y^a | S = 1]$, and non-randomized individuals, $\widehat  \E[Y^a | S = 0]$, is of practical interest: provided the identifiability conditions hold, if $\widehat \E[Y^a | S = 1] \approx \widehat  \E[Y^a | S = 0]$, we may conclude that the trial results are likely generalizable (up to sampling variability); in contrast, if the estimates are different, trial results may not be generalizable.

We also showed that the different study designs have implications for identifying and estimating the conditional probability of trial participation. This probability is of inherent interest because it captures aspects of decision-making related to trial participation \cite{mcfadden1973conditional, Stuart2011}. We showed that the probability is identifiable in nested trial designs, but not in non-nested trial designs (e.g., composite dataset designs). Indeed, any reasonable parametric model for the probability of participation in the population can be identified in nested trial designs. In nested trial designs with sampling of non-randomized individuals, estimation of model parameters can be facilitated by the use of weighted maximum likelihood estimation where randomized patients are given weight 1 and non-randomized patients are given weight equal to the inverse of the probability of being sampled among non-randomized individuals in the actual population. In non-nested trial designs, model specification is complicated by the fact that, when sampling depends on trial participation status, the model for the probability of trial participation among sampled individuals is not of the same form as the model in the population (the logistic regression model being a notable exception \cite{manski1977estimation}).

The probability of trial participation in the target population is also important for identification and estimation using inverse probability (or odds) weighting methods. Our argument about the odds of participation after selection of non-randomized individuals being equal to the odds of participation in the target population up to an unknown multiplicative constant, clarifies how the validity of estimators when using composite datasets designs \cite{westreich2017, dahabreh2018transporting} depends critically on the assumed sampling properties.

Astute readers will have noticed the many connections between our results and the theory of case-referent (case-control) studies \cite{miettinen1976estimability, mantel1973synthetic,manski1977estimation, cosslett1981maximum, scott1986fitting}. Indeed, our approach can be placed in the case-base paradigm, viewing randomized individuals as ``cases'' in cumulative incidence case-referent study \cite{miettinen1976estimability} nested in the ``cohort'' of the actual population. An interesting parallel with case-referent studies: the difficulty in specifying the population of non-randomized individuals that should be sampled in composite dataset designs is similar in nature to the validity issues of case-referent studies with a secondary base \cite{miettinen1985case, miettinen1990response, wacholder1992selection}.

In this paper, for simplicity, we focused on causal quantities that are most meaningful for point treatments with complete adherence and no loss to follow-up. Our results can be extended to address time-varying treatments using well-known extensions of the identifiability conditions for randomized trials \cite{robins1986, robins2000b, hernan2019}, without any changes to the sampling properties or the modeling assumptions about the probability of trial participation. Perhaps, then, the most consequential causal assumption that we required was that the invitation to participate in the trial and trial participation itself do not have an effect on the outcome except through treatment assignment. Unless investigators are willing to contemplate much more complex study designs involving multistage data collection about (and possibly randomization of) the invitation to participate, trial participation itself, and treatment assignment \cite{heckman1991randomization}, our results are best viewed as applying to trials where the not-through-treatment effects of the invitation to participate in the trial and of trial participation are negligible compared to the effects of treatment. For example, they are applicable to pragmatic randomized trials embedded in large health-care systems or registries, where trial procedures other than treatment assignment can be assumed to be similar to usual medical practice \cite{staa2014a, ford2016, choudhry2017}.

\clearpage
\section{Figure}

\vspace{0.4in}

\clearpage

\begin{landscape}

\section{Table}
{
\renewcommand{\arraystretch}{2}
\begin{table}[!ht]
\caption{Summary of sampling properties and identification results by study design.}
\label{tab:summary_conditions_identification}
\resizebox{1.4\textwidth}{!}{%
\begin{tabular}{|l|l|l|l|l|}
\hline
\textbf{Study design}             & \textbf{Sampling probabilities}                                                                                                           & \textbf{\begin{tabular}[c]{@{}l@{}}Marginal probability \\ of trial participation\end{tabular}}       & \textbf{\begin{tabular}[c]{@{}l@{}}Conditional probability \\ of trial participation\end{tabular}}              & \textbf{\begin{tabular}[c]{@{}l@{}}Identifiable potential \\ outcome means \\ (when the conditions \\ in Section \ref{sec:identifiability_conditions} hold)\end{tabular}} \\ \hline
\multirow{2}{*}{Nested-trial}                      & \begin{tabular}[c]{@{}l@{}}$\Pr[D = 1 | S = 1] = 1$ and \\ $\Pr[D = 1 | S = 0] = 1$\end{tabular}                                          & $\Pr[S = 1] = \left\{ 1 + \dfrac{\Pr[S = 0 | D = 1]}{\Pr[S = 1 | D = 1]} \right\}^{-1}$               & $\Pr[S = 1 | X] = \Pr[S = 1 | X, D = 1]  $                                                                      & $\E[Y^a]$ and $\E[Y^a | S = 0]$   \\
\cline{2-5}  
 & \begin{tabular}[c]{@{}l@{}}$\Pr[D = 1 | S = 1] = 1$ and \\ $\Pr[D = 1 | X, A, Y, S = 0] = \Pr[D = 1 | S = 0] = c > 0$, \\ $c$ is a known constant\end{tabular}    & $ \Pr[S = 1] = \left\{1 + \dfrac{\Pr[S = 0 | D = 1]}{\Pr[S = 1 | D = 1]} \times c^{-1} \right\}^{-1}$ & $ \Pr[S = 1 | X] = \left\{ 1 + \dfrac{\Pr[S=0 | X , D = 1]}{\Pr[S = 1 | X, D = 1]} \times c^{-1} \right\}^{-1}$ & $\E[Y^a]$ and $\E[Y^a | S = 0]$                                                                                                         \\ \hline
      Non-nested trial      & \begin{tabular}[c]{@{}l@{}}$\Pr[D = 1 | S = 1] = 1$ and \\ $\Pr[D = 1 | X, A, Y, S = 0] =\Pr[D = 1 | S = 0] = u > 0,$ \\ $u$ is an unknown constant\end{tabular} & Not identifiable                                                                                      & Not identifiable                                                                                                & $\E[Y^a | S = 0]$                                                                                                                            \\ \hline
\end{tabular}
}
\caption*{Note that the formulas for the marginal and conditional probability of trial participation in nested-trial designs with a census of the actual population can be obtained from the formulas for the nested trial design with known sampling probabilities by setting $c=1$.}
\end{table}
} 

\end{landscape}

\clearpage
\bibliographystyle{ieeetr}
\bibliography{transportability_study_design}

\clearpage
\appendix

\section{Identification of the probability of trial participation in nested trial designs with sub-sampling}

\setcounter{equation}{0}
\renewcommand{\theequation}{A.\arabic{equation}}

\subsection{Identification of the marginal probability of trial participation}

Using the definition of conditional probability and re-arranging, $$ \Pr[S = 1 ] = \dfrac{\Pr[S = 1 | D = 1] \Pr[ D = 1 ]}{\Pr[D = 1 | S = 1]} \mbox{ and } \Pr[S = 0 ] = \dfrac{\Pr[S = 0 | D = 1] \Pr[ D = 1 ]}{\Pr[D = 1 | S = 0]}.$$ Taking the ratio of the above expressions and exploiting the sampling properties for non-nested trial designs, 
\begin{equation*} 
	\begin{split}
\dfrac{\Pr[S = 1 ]}{\Pr[S = 0]} &= \dfrac{\Pr[S = 1 | D = 1]}{\Pr[S = 0 | D = 1]} \times  \dfrac{\Pr[D = 1 | S = 0]}{ \Pr[D = 1 | S = 1]} \\[10pt]
	&= \dfrac{\Pr[S = 1 | D = 1]}{\Pr[S = 0 | D = 1]} \times c.
	\end{split}
\end{equation*}
With a bit of algebra, the above expression can be re-arranged to show that
\begin{equation*} 
\Pr[S = 1 ] = \left\{ 1 + \dfrac{\Pr[S = 0 | D = 1]}{\Pr[S = 1 | D = 1]} \times  c^{-1} \right\}^{-1}.
\end{equation*}

By setting $c=1$ we see that in the nested-trial design with a census of non-randomized individuals $\Pr[S = 1] = \Pr[S = 1 | D = 1]$.

\subsection{Identification of the conditional probability of trial participation}

The argument for the conditional probability is parallel to the one presented above for the marginal probability. Again, using the definition of conditional probability,
\begin{equation*} 
	\setlength{\jot}{10pt}
	\begin{split}
\Pr[S = 1 | X ] &= \dfrac{\Pr[S = 1 | X, D = 1] \Pr[ D = 1 | X ]}{\Pr[D = 1 | X, S = 1]} \mbox{ and } \\[10pt] 
 \Pr[S = 0 | X ] &= \dfrac{\Pr[S = 0 | X, D = 1] \Pr[ D = 1 | X ]}{\Pr[D = 1 | X, S = 0]}.
	\end{split}
\end{equation*} 
Taking the ratio of the above expressions and exploiting the sampling properties for non-nested trial designs, 
\begin{equation*} 
	\begin{split}
\dfrac{\Pr[S = 1 |X]}{\Pr[S = 0 | X]} &= \dfrac{\Pr[S = 1 | X, D = 1]}{\Pr[S = 0 | X, D = 1]} \times  \dfrac{\Pr[D = 1 | X , S = 0]}{ \Pr[D = 1 | X, S = 1]} \\[10pt]
	&= \dfrac{\Pr[S = 1 | X, D = 1]}{\Pr[S = 0 | X,  D = 1]} \times c.
	\end{split}
\end{equation*}
The above expression can be re-arranged to show that 
\begin{equation*}
	\begin{split}
\Pr[S = 1 | X] =  \left\{ 1 + \dfrac{\Pr[S=0 | X , D = 1]}{\Pr[S = 1 | X, D = 1]} \times c^{-1} \right\}^{-1}.
	\end{split}
\end{equation*}

By setting $c=1$ we see that in the nested-trial design with a census of non-randomized individuals $\Pr[S = 1| X ] = \Pr[S = 1 | X, D = 1]$.

\clearpage
\section{Estimating the conditional probability of trial participation}

\setcounter{equation}{0}
\renewcommand{\theequation}{B.\arabic{equation}}

We outline the proof for the convergence in probability of the estimators for the conditional probability of trial participation described in the main text, without delving into the technical conditions needed to make the arguments rigorous. 

Consider the likelihood function for the nested trial design with a census of the actual population,
\begin{equation*}
  \begin{split}
  \mathscr{L}(\gamma) &= \prod\limits_{i=1}^{n} \big[ p(X_i; \gamma)\big]^{S_i} \big[ 1 - p(X_i; \gamma)\big]^{1 - S_i},
  \end{split}
\end{equation*}
and, the pseudo-likelihood function for the nested trial design with known sampling probability of the non-randomized individuals,
\begin{equation*}
  \begin{split}
  \mathscr{L}_{\scalebox{.5}{\text{W}}}(\gamma) &= \prod\limits_{i=1}^{n} \big[ p(X_i; \gamma)\big]^{S_iD_i} \big[ 1 - p(X_i; \gamma)\big]^{[(1 - S_i) D_i ] / c}.
  \end{split}
\end{equation*}

For $\mathscr{L}(\gamma)$, the sample size-normalized objective function to be maximized is 
\begin{equation*}
  \begin{split}
  \widehat{\ell}(\gamma) &= \dfrac{1}{n}\sum\limits_{i=1}^{n} \Big\{ S_i  \log p(X_i; \gamma) +  (1 - S_i) \log \big[  1 - p(X_i; \gamma) \big]  \Big\}.
  \end{split}
\end{equation*}
Provided the technical conditions for the uniform law of large numbers obtain, the above objective function converges uniformly in probability, in the sense of the definition in Section 2.1 of \cite{newey1994large}, to 
\begin{equation*}
  \ell_0(\gamma) = \E\! \Big[ S \log  p(X; \gamma)  + (1 - S) \log \big[1 - p(X; \gamma)\big] \Big].
\end{equation*}
By Theorem 2.1 of \cite{newey1994large}, if $\ell_0(\gamma)$ is uniquely maximized at $\gamma_0$, the parameter space is compact, and $\ell_0(\gamma)$ is continuous, then the estimator $\widehat \gamma$ obtained by maximizing $\widehat \ell(\gamma)$, converges in probability to $\gamma_0$, that is, $\widehat \gamma \overset{p}{\longrightarrow} \gamma_0$.

For $\mathscr{L}_{\scalebox{.5}{\text{W}}}(\gamma)$, the sample size-normalized objective function to be maximized is 
\begin{equation*}
  \widehat{\ell}_{\scalebox{.5}{\text{W}}}(\gamma) = \dfrac{1}{n}\sum\limits_{i=1}^{n} \left\{ S_i D_i  \log p(X_i; \gamma) + \dfrac{(1 - S_i) D_i}{c} \log \big[  1 - p(X_i; \gamma) \big] \right\}.
\end{equation*}
Because $c$ is assumed to be bounded away from 0, and provided the technical conditions for the uniform weak law of large numbers obtain, the above objective function converges uniformly in probability to
\begin{equation*}
  \ell_{\scalebox{.5}{\text{W}}0}(\gamma) = \E\! \left[ S D \log  p(X; \gamma)  + \dfrac{(1 - S) D}{c} \log \big[1 - p(X; \gamma)\big] \right].
\end{equation*}

We will now show that $\ell_0(\gamma) =  \ell_{\scalebox{.5}{\text{W}}0}(\gamma)$. 

By design, if $S =1 $, then $ S D = 1$; if $S = 0$, then $S   D = 0$. Thus, to establish the result we only need to show that $$\E \big[ (1 - S) \log [  1 - p(X; \gamma) \big] = \E \left[ \dfrac{(1 - S) D}{c} \log \big[  1 - p(X; \gamma) \big] \right].$$ Starting from the right-hand-side, 
\begin{equation*}
  \begin{split}
   \E \left[ \dfrac{(1 - S) D}{c} \log \big[  1 - p(X; \gamma) \big] \right] &= \E \Big[ \log \big[  1 - p(X; \gamma) \big] \big | S = 0, D = 1 \Big] \dfrac{\Pr[S = 0, D = 1 ]}{c} \\ 
   &= \E \Big[ \log \big[  1 - p(X; \gamma) \big] \big | S = 0 \Big] \Pr[S = 0 ] \\
   &= \E \Big[ (1 - S) \log \big[  1 - p(X; \gamma) \big] \Big],
  \end{split}
\end{equation*}
which establishes the result.

Because $\ell_0(\gamma) = \ell_{\scalebox{.5}{\text{W}}0}(\gamma)$, it follows that the maximizer of $\widehat \ell_{\scalebox{.5}{\text{W}}}(\gamma)$, $\widehat{\gamma}_{\scalebox{.5}{\text{W}}}$, converges in probability to $\gamma_0$, that is,  $\widehat{\gamma}_{\scalebox{.5}{\text{W}}} \overset{p}{\longrightarrow} \gamma_0$.

To obtain the asymptotic distribution of the estimators, we need additional technical conditions as in Theorem 3.1 of \cite{newey1994large}; provided these conditions hold, $\widehat \gamma$ and $\widehat{\gamma}_{\scalebox{.5}{\text{W}}}$ are asymptotically normal.

\clearpage
\section{Nested trial design with covariate-dependent sampling probabilities} 

\setcounter{equation}{0}
\renewcommand{\theequation}{B.\arabic{equation}}

\subsection{Sampling properties}

As noted in the main text, a more general version of the nested trial design assumes that the sampling probabilities for non-randomized individuals depend on baseline auxiliary covariates. Let $X = (X_1, X_2)$, where $X_1$ represents baseline auxiliary covariates that are available on all members of the actual population in which the trial is nested, and $X_2$ represents covariates that are only measured among randomized individuals ($S=1$) and sampled non-randomized individuals ($S = 1, D=1$). 

The identifiability conditions and identification results remain the same as in the main text; but the sampling properties of this design are 
\begin{equation*} \label{eq:sampling_ass_2sample_aux}
	\begin{split}
		&\Pr[D = 1 | S = 1] = 1 \mbox{, and } \\
    &\Pr[D = 1 | X, A, Y, S = 0] = \Pr[D = 1 | X_1, S = 0] \equiv c(X_1), 
	\end{split}
\end{equation*}
where $0 < c(X_1) \leq 1$ is a known function that only depends on $X_1$, allowing the sampling of non-randomized individuals to depend on the auxiliary covariates that are available from all members of the actual population.

\subsection{Identification of the conditional probability of trial participation}

Using an argument similar to the case when the sampling probability for non-randomized individuals does not depend on covariates, we obtain 
\begin{equation*}
	\begin{split}
\Pr[S = 1 | X] =  \left\{ 1 + \dfrac{\Pr[S=0 | X , D = 1]}{\Pr[S = 1 | X, D = 1]} \times \dfrac{1}{c(X_1)} \right\}^{-1},
	\end{split}
\end{equation*}
which is identifiable because the inverse of the conditional odds of trial participation in the sampled data, $\dfrac{\Pr[S=0 | X , D = 1]}{\Pr[S = 1 | X, D = 1]}$, are identifiable, and $c(X_1)$ is known, by design.

\subsection{Estimating the probability of trial participation by weighted regression}

As before, we assume a model $p(X; \gamma)$ for $\Pr[S = 1 | X]$ with finite-dimensional parameter $\gamma$. The weighted pseudo-likelihood function becomes
\begin{equation*}
  \begin{split}
  \mathscr{L}^{*}_{\scalebox{.5}{\text{W}}}(\gamma) &= \prod\limits_{i=1}^{n} \big[ p(X_i; \gamma)\big]^{S_iD_i} \big[ 1 - p(X_i; \gamma)\big]^{[ (1 - S_i) D_i ] / c(X_{1i})}.
  \end{split}
\end{equation*}
Note that the only difference between $\mathscr{L}^{*}_{\scalebox{.5}{\text{W}}}(\gamma)$ and $\mathscr{L}_{\scalebox{.5}{\text{W}}}(\gamma)$ is that the weights in the former depend on $X_1$. 
The sample size-normalized objective function to be maximized is 
\begin{equation*}
  \widehat{\ell}^*_{\scalebox{.5}{\text{W}}}(\gamma) = \dfrac{1}{n}\sum\limits_{i=1}^{n} \left\{ S_i D_i  \log p(X_i; \gamma) + \dfrac{(1 - S_i) D_i}{c(X_{1i})} \log \big[  1 - p(X_i; \gamma) \big] \right\}.
\end{equation*}
Because $c(X_1)$ is assumed to be bounded away from 0, and provided the technical conditions for the uniform weak law of large numbers obtain, the above objective function converges uniformly in probability to
\begin{equation*}
  \ell^{*}_{\scalebox{.5}{\text{W}}0}(\gamma) = \E\! \left[ S D \log  p(X; \gamma)  + \dfrac{(1 - S) D}{c(X_1)} \log \big[1 - p(X; \gamma)\big] \right].
\end{equation*}


We will now show that $\ell_0(\gamma) =  \ell^*_{\scalebox{.5}{\text{W}}0}(\gamma)$.

As noted above, by design, if $S =1 $, then $ S D = 1$; if $S = 0$, then $S   D = 0$. Thus, to establish the result we only need to show that $$\E \Big[ (1 - S) \log \big[  1 - p(X; \gamma) \big] \Big] = \E \left[ \dfrac{(1 - S) D}{c(X_1)} \log \big[  1 - p(X; \gamma) \big] \right].$$ Starting from the right-hand-side, 
\begin{equation*}
  \begin{split}
   \E \left[ \dfrac{(1 - S) D}{c(X_1)} \log \big[  1 - p(X; \gamma) \big] \right] &= \E \Bigg[ \E \left[ \dfrac{(1 - S) D}{c(X_1)} \log \big[  1 - p(X; \gamma) \big] \Big| X_1 \right]  \Bigg] \\
   &= \E \Bigg[ \dfrac{1}{c(X_1)} \E \left[ (1 - S) D \log \big[  1 - p(X; \gamma) \big] \big| X_1 \right]  \Bigg] \\
   &= \E \Bigg[ \dfrac{\Pr[S = 0, D = 1 | X_1]}{c(X_1)} \E \left[ \log \big[  1 - p(X; \gamma) \big] \big| X_1, S = 0, D = 1 \right]  \Bigg] \\
   &= \E \Bigg[ \Pr[S = 0 | X_1] \E \left[ \log \big[  1 - p(X; \gamma) \big] \big| X_1, S = 0, D = 1 \right]  \Bigg] \\
   &= \E \Bigg[ \Pr[S = 0 | X_1] \E \left[ \log \big[  1 - p(X; \gamma) \big] \big| X_1, S = 0 \right]  \Bigg] \\
   &= \E \Bigg[ \E \left[ (1 - S) \log \big[  1 - p(X; \gamma) \big] \big| X_1 \right]  \Bigg] \\
  &= \E \Big[ (1 - S) \log \big[  1 - p(X; \gamma) \big] \Big],   \\
  \end{split}
\end{equation*}
which establishes the result.

Because $\ell_0(\gamma) = \ell^{*}_{\scalebox{.5}{\text{W}}0}(\gamma)$, it follows that the maximizer of $\widehat \ell^{*}_{\scalebox{.5}{\text{W}}}(\gamma)$, $\widehat{\gamma}^{*}_{\scalebox{.5}{\text{W}}}$, converges in probability to $\gamma_0$, that is,  $\widehat{\gamma}^{*}_{\scalebox{.5}{\text{W}}} \overset{p}{\longrightarrow} \gamma_0$.

In practical terms, this result suggests that the conditional probability of trial participation in the target population can be estimated using a weighted regression of $S$ on $X$ among sampled patients, using weights equal to 1 for randomized patients (all of whom are sampled); $1/c(X_1)$ for sampled non-randomized individuals; $0$ for non-sampled non-randomized individuals. 

As above, provided the technical conditions of Theorem 3.1 of \cite{newey1994large} hold, $\widehat{\gamma}^{*}_{\scalebox{.5}{\text{W}}}$ is asymptotically normal.


\ddmmyyyydate 
\newtimeformat{24h60m60s}{\twodigit{\THEHOUR}.\twodigit{\THEMINUTE}.32}
\settimeformat{24h60m60s}
\begin{center}
\vspace{\fill}\ \newline
\textcolor{black}{{\tiny $ $transportability\_study\_design, $ $ }
{\tiny $ $Date: \today~~ \currenttime $ $ }
{\tiny $ $Revision: \paperversionmajor.\paperversionminor $ $ }}
\end{center}

\end{document}